\documentclass[12pt,noshowpacs,nofootinbib,notitlepage,amsmath,amssymb]{revtex4-2}
\usepackage{setspace}
\usepackage[top=1in,bottom=1in,left=1in,right=1in]{geometry}
\usepackage{graphicx,color,enumitem}
\usepackage[colorlinks=true,citecolor=blue,linkcolor=blue,urlcolor=blue]{hyperref}

\newcommand{\comm}[2]{[#1,#2]}
\newcommand{\acomm}[2]{\{#1,#2\}}

\newcommand{\PV}{\mathcal{P}}

\newcommand{\ket}[1]{|#1\rangle}
\newcommand{\bra}[1]{\langle#1|}

\begin{document}
\title{\color{blue}\Large Superselected ghost theory: perturbation theory}
\author{Bob Holdom}
\email{bob.holdom@utoronto.ca}
\affiliation{Department of Physics, University of Toronto\\ Toronto, Ontario, Canada  M5S 1A7}

\begin{abstract}
A superselection rule based on an exact ghost parity can endow a ghost QFT with a probability interpretation. However, this ghost parity is not respected at finite order in the standard perturbative expansion. A ghost-parity-preserving perturbation theory (Z$_2$PT) is obtained through a similarity transformation of the Hamiltonian, $h=g H g^{-1}=h_0+h_1+h_2+...$. The resulting expansion is reminiscent of old-fashioned perturbation theory (OFPT) with some significant differences. The superselection rule selects the principal-value prescription for cross-sector energy denominators, with the prescription determined by the type of transition rather than the particle species. At third order, products of $h_1$ and $h_2$ combine with $h_3$ to reproduce OFPT away from vanishing denominators. At fourth order, when $h_1=0$, we show that $h_2^2$ and $h_4$ satisfy the analogous relation. Contact terms from vanishing denominators distinguish Z$_2$PT from OFPT but do not alter the local primitive UV divergences through these orders.
\end{abstract}

\maketitle

\section{Introduction}
Consider a ghost QFT in the real-spectrum regime subject to a superselection rule; the exact
Hamiltonian $H$ commutes with a ghost-parity operator $Q$ satisfying $Q^2=1$. This yields a theory with a probability interpretation \cite{paper1}. By contrast, the perturbative splitting $H=H_0+H_1$ does not
respect this symmetry; in general, $\comm{Q}{H_0}\neq0$ and $\comm{Q}{H_1}\neq0$. The free part $H_0$ is
$\eta$-preserving ($\comm{\eta}{H_0}=0$), where $\eta$ defines the native indefinite inner product and satisfies $\eta^2=1$.

With respect to the native inner product, $H$ and $Q$ are $\eta$-pseudo-Hermitian:
$H^\dagger\eta=\eta H$ and $Q^\dagger\eta=\eta Q$. In the real-spectrum regime, there also exists a positive-definite inner product defined by $G=\eta Q$. Both $H$ and $Q$ are pseudo-Hermitian with respect to $G$ as well, $H^\dagger G=GH$ and $Q^\dagger G=GQ$. Of particular importance here, $g=\sqrt{G}$ defines the similarity transformation $h=gHg^{-1}$. Through this transformation, perturbation theory can be reorganized so that ghost parity, represented by $\eta$, is manifest at every order.

Section~\ref{sec:general} develops the
general framework of ghost-parity-preserving perturbation theory (Z$_2$PT), deriving the similarity
transformation $h=gHg^{-1}$ order by order and obtaining compact expressions
for $h_1$ through $h_4$. Section~\ref{sec:matrix} evaluates matrix elements
in the Fock basis, where the superselection rule selects the principal-value
prescription for cross-sector energy denominators, and shows that $h_2$
reproduces the real part of the Feynman amplitude. Section~\ref{sec:h3structure}
examines the new structural features that appear at third order; mixed
prescriptions and the organization of
perturbation theory by the sector-transition pattern. Section~\ref{sec:h4structure}
studies the fourth-order terms $h_2^2$ and $h_4$ when $h_1=0$ and compares
their energy-denominator and local UV structure with OFPT. At these orders and away from
vanishing denominators, the
nonstandard energy denominators that appear in individual terms do not appear when summed.
Section~\ref{sec:renormalizability} comments on renormalization and the
local $\mu$-running of Z$_2$PT couplings.

The similarity transformation in Z$_2$PT plays a role analogous to those of two other constructions.
In pseudo-Hermitian / PT-symmetric quantum mechanics,
a non-Hermitian Hamiltonian $H$ with real spectrum is mapped to a
Hermitian Hamiltonian \cite{Mostafazadeh:2001nr}
(for reviews, see \cite{Bender:2007review,Mostafazadeh:2010review}). The Schrieffer--Wolff (SW) transformation is used in
condensed matter physics to eliminate couplings between low- and
high-energy subspaces~\cite{sw_h3,Bravyi:2011fda}. In Section~\ref{sec:comparison}, we contrast Z$_2$PT with the perturbative expansions used in these approaches. We conclude in Section~\ref{conc}.

\section{Transformed perturbation theory}\label{sec:general}

\subsection{Setup}

The similarity transformation can be constructed perturbatively as follows,
\begin{align}
  g &= \sqrt{G},\qquad   G = \eta Q,\\
  Q &= \eta + Q_1+Q_2+Q_3+\cdots,\\
  h &= g H g^{-1} = h_0 + h_1 + h_2 + h_3 + \cdots,\\
  \eta &= g Q g^{-1}.\label{e3}
\end{align}
Under this transformation, $\comm{Q}{H}=0$ becomes $\comm{\eta}{h}=0$, so $\eta$
is the preserved ghost parity in the transformed theory. Thus, $\eta$ serves both as the inner product and as the exact ghost parity. The transformed Hamiltonian is Hermitian, $h=h^\dagger$, and Eq.~(\ref{e3}) is derived in \cite{paper1}.
Both $h_n$ and $Q_n$ involve $n$ insertions of $H_1$. Thus, the expansion of $h$ is a coupling expansion; for example, $h_2$ can contribute at both tree level and one loop.

The commutation relations for $Q_n$ implied by $\comm{Q}{H}=0$ are
\begin{align}
  \comm{Q_1}{H_0} &= -\comm{\eta}{H_1}, \label{eq:Q1comm}\\
  \comm{Q_2}{H_0} &= -\comm{Q_1}{H_1}, \\
  \comm{Q_n}{H_0} &= -\comm{Q_{n-1}}{H_1}, \qquad n\ge 2,
\end{align}
and the conditions implied by $Q^2=1$ are
\begin{align}
  \acomm{\eta}{Q_1} &= 0, \label{eq:grade1}\\
  \acomm{\eta}{Q_2} &= -Q_1^2, \\
  \acomm{\eta}{Q_n} &= -\sum_{i=1}^{n-1} Q_i Q_{n-i}, \qquad n\ge 2.
\end{align}

Every operator can be decomposed as $A = A^\parallel + A^\perp$, with
\begin{equation}\label{eq:projectors}
  A^\parallel = \tfrac12(A + \eta A\eta),\qquad
  A^\perp   = \tfrac12(A - \eta A\eta),
\end{equation}
where $A^\parallel$ is $\eta$-preserving, $\comm{\eta}{A^\parallel}=0$, and $A^\perp$ is $\eta$-flipping, $\acomm{\eta}{A^\perp}=0$.
We see that $Q_1$ is $\eta$-flipping, while $Q_n$ for $n\ge2$ generally contains both
$\eta$-preserving and $\eta$-flipping parts.

\subsection{Leading orders}

The raw expansion of $gHg^{-1}$ to order $n$ produces terms
containing $H_0$, $H_1$, $\eta$, and $Q_1,\ldots,Q_n$.
The commutation rules allow $H_0$ to be pushed to the left through any $Q_k$
and thus moved to the left edge of each term. The resulting
$H_0$-dependent terms cancel, leaving an expression involving only
$H_1$, $\eta$, and the $Q_k$. This expression can be further simplified
using the $Q^2=1$ relations.

It is convenient to define
\begin{equation}
  J =\comm{\eta}{H_1}=\comm{\eta}{H}=\comm{\eta}{H^\perp}
  =2\eta H^\perp.
\end{equation}
The pseudo-Hermiticity and $\eta$-(anti)commutation properties imply that
$J$ and $H_1^\parallel$ are Hermitian, while $H^\perp$, $Q_1$, and
$Q_n^\perp$ are anti-Hermitian; for example,
$H^{\perp\dagger}=-H^\perp$.

The transformed Hamiltonian through the fourth order is
\begin{align}
  h_0 &= H_0, \\
  h_1 &= \tfrac12 H_1 + \tfrac12 \eta H_1\eta
  = H_1^\parallel, \label{eq:h1def}\\
  h_2 &= \tfrac18\comm{J}{Q_1},
  \label{eq:h2def}\\
  h_3 &= \tfrac18\comm{J}{Q_2^\perp}, \\
  h_4 &= \tfrac18\comm{J}{Q_3^\perp}
  + \tfrac3{128}\comm{J}{Q_1^3}
  + \tfrac1{128}\,Q_1\comm{Q_1}{J}Q_1. \label{eq:simplified}
\end{align}
$[\eta,h_i]=0$, and the same holds for each of the three terms in $h_4$.
Since $h_i$ involves $i$ insertions of $H_1$, if
$h_1=H_1^\parallel=0$, then $h_i$ vanishes for every odd $i$, because
odd powers of $H^\perp$ cannot be $\eta$-preserving.

In the next section, we show how the $Q_i$ depend on energy differences,
which are in turn determined by spatial momenta. The $Q_i$ are therefore
nonlocal in space, implying that the operators $h_i$ for $i>1$ are spatially nonlocal
while local in time. They encode virtual transitions across the
$\eta$-gap. In contrast, on-shell intermediate states that respect the
superselection rule are generated by time-ordered products of the $h_i$.

\section{Matrix elements}\label{sec:matrix}

With respect to the $\eta$ inner product, matrix elements are of the form $\langle m|\eta\cdots|n\rangle$. We consider matrix elements between Fock
states $\ket{m}$ and $\ket{n}$ in a definite $\eta$-sector. At first order, we have the simple local operator $h_1 = H_1^\parallel$. At second order, $h_2$ encodes transitions to and from the opposite $\eta$-sector. We insert a complete set of intermediate states $\ket{a}$, all of which must belong to that sector because $J$ and $Q_1$ are $\eta$-flipping,
\begin{equation}\label{eq:h2expand}
  \bra{m}\eta h_2\ket{n}
  = \frac18\sum_a \frac{\bra{m}\eta J\ket{a}\bra{a}\eta Q_1\ket{n}}{\bra{a}\eta\ket{a}}
  - \frac18\sum_a \frac{\bra{m}\eta Q_1\ket{a}\bra{a}\eta J\ket{n}}{\bra{a}\eta\ket{a}}.
\end{equation}
The sum includes 3-momentum integrals. The action of $\eta$ gives a definite sign determined by the state. The net result, after letting the sum absorb the standard positive $1/\langle a|a\rangle$ factors, is
\begin{equation}\label{eq:h2expand2}
  \bra{m}h_2\ket{n}
  = \frac18\sum_a \bra{m}J\ket{a}\bra{a}Q_1\ket{n}
  - \frac18\sum_a \bra{m}Q_1\ket{a}\bra{a}J\ket{n}.
\end{equation}
This holds whether the states $\ket{m}$ and $\ket{n}$ or the states $\ket{a}$ are in the $\eta=-1$ sector. Thus, we can effectively calculate matrix elements as if the inner product were positive definite, consistent with the hermiticity of $h$. We shall reduce these expressions to matrix elements of $H^\perp$, where the ghost sign is encoded through the anti-Hermiticity of $H^\perp$.

To determine the matrix elements of $Q_1$, we start from Eq.~(\ref{eq:Q1comm}), $\comm{Q_1}{H_0}=-J$. In the Fock basis,
where $H_0\ket{n}=E_n\ket{n}$ with $E_n$ determined by spatial momenta, they satisfy
\begin{equation}\label{eq:Q1bare}
  \bra{a}Q_1\ket{n} = \frac{\bra{a}J\ket{n}}{E_a-E_n}
  \qquad (E_a\neq E_n).
\end{equation}
The inverse energy denominator is understood distributionally in the
continuum limit. In standard OFPT,
\begin{equation}
  \frac{1}{x+i0}
  = \PV\frac{1}{x}-i\pi\delta(x).
\end{equation}
The delta-function term represents an on-shell transition, but an on-shell transition
between states of opposite $\eta$ sectors is forbidden by the superselection rule.
We therefore retain the symmetric real boundary value,
\begin{equation}\label{eq:Q1prescriptions}
  \bra{a}Q_1\ket{n}
  = \frac12\lim_{\varepsilon\to0^+}
    \left(\frac{1}{E_a-E_n+i\varepsilon}
         +\frac{1}{E_a-E_n-i\varepsilon}\right)
    \bra{a}J\ket{n}
  = \PV\frac{\bra{a}J\ket{n}}{E_a-E_n}.
\end{equation}
$\PV$ does not assign an ordinary value at exact degeneracy; it defines
the momentum integrals across the singularity.

Using this prescription and $J=2\eta H^\perp$, with $\eta$ again producing signs determined by the adjacent state, we obtain
\begin{align}
  \bra{m}h_2\ket{n}
  &= \frac12\sum_a\bra{m}H^\perp\ket{a}\bra{a}H^\perp\ket{n}
  \PV\left[\frac{1}{E_n-E_a}+\frac{1}{E_m-E_a}\right]\label{eq:h2offdiag}\\
  &= -\sum_a\bra{m}H^{\perp\dagger}\ket{a}\bra{a}H^\perp\ket{n}
  \PV\frac{1}{E_n-E_a}\quad\mbox{for } E_m=E_n,\\
  &= -\sum_a\PV\frac{|\bra{a}H^\perp\ket{n}|^2}{E_n-E_a}\quad\mbox{for } m=n.\label{e1}
\end{align}
In the second equality, we used $H^{\perp\dagger}=-H^\perp$, manifesting the minus sign intrinsic to ghost theories.

The sum over $a$ includes two classes of intermediate states. The first consists of states that can produce vanishing denominators, which are handled using $\PV$. The second consists of $Z$-diagram states, for which $E_a$ is always
separated from $E_n$ (typically $E_a>E_n$ because additional particles
are present), so the denominator never vanishes. These states contribute a smooth real background included in the standard Feynman amplitude.

Except for the replacement of the $i\varepsilon$ prescription by $\PV$, Eq.~(\ref{e1}) is the result expected from old-fashioned perturbation theory (OFPT)~\cite{ofpt} applied to a ghost theory with an anti-Hermitian $H^\perp$. OFPT and Feynman rules give equivalent results when applied to the same ghost theory. For physical transitions with $E_m=E_n$, we therefore have
\begin{align}\label{eq:h2equalsReFeynman}
  \bra{m}h_2\ket{n} &=\operatorname{Re}\bigl[\text{OFPT with two $H^\perp$ insertions}\bigr]\nonumber\\
  &= \operatorname{Re}\,
  \bigl[\text{Feynman diagram with two $H^\perp$ insertions}\bigr].
\end{align}
This applies to second-order results, including both tree-level diagrams and one-loop self-energy diagrams.\footnote{In the Feynman calculation of the self-energy, the relevant quantity is the relative sign between the Feynman self-energy and the propagator being corrected.} This $\operatorname{Re}[\text{Feynman}]$ equivalence does not hold at higher orders.

$h_2$ shows how an interaction $H^\perp$ that would violate the superselection rule in the original perturbation theory is repackaged. Such interactions still induce virtual transitions to and from the opposite sector, but without any superselection-rule-violating manifestations. At second order, there are also matrix elements of time-ordered products, $\bra{m}T[h_1(t_1)h_1(t_2)]\ket{n}$, which give the standard Feynman or OFPT results for two insertions of $H_1^\parallel$. In what follows, it will be more convenient to compare directly with OFPT expressions.

\section{Structure at third order}\label{sec:h3structure}

At third order, contributions arise from three terms in the Dyson expansion
of the $S$-matrix, $T[h_1(t_1)h_1(t_2)h_1(t_3)]$,
$T[h_1(t_1)h_2(t_2)]$, and $h_3(t_1)$. The first reproduces standard
OFPT at cubic order in the interaction $H_1^\parallel$; the other two contain two $H^\perp$
insertions and one $H_1^\parallel$ insertion and do not reproduce OFPT.

\subsection{The $h_1 \cdot h_2$ mixed term}
\label{sec:h1h2}

$T[h_1(t_1)h_2(t_2)]$ produces two time orderings, each with a
denominator $1/(E_i-E_a+i\varepsilon)$, where $\ket{a}$ is the
intermediate state between the two insertions. Since all $h_i$ are
sector-preserving, $\ket{a}$ is in the same $\eta$-sector as the
initial state, which we take, for definiteness, to have $\eta=1$.
Meanwhile, the states $\ket{b}$ are the $\eta=-1$ intermediate states
within $h_2$, and their energy denominators carry the $\PV$ prescription.
The two time orderings of $h_1$ and $h_2$ give
\begin{align}
  \bra{f}h_1\!\cdot\! h_2\ket{i}
  &= \sum_{a\in\eta=+1}\sum_{b\in\eta=-1}
  \frac{\bra{f}H_1^\parallel\ket{a}}{E_i-E_a+i\varepsilon}\;
  \bra{a}H^\perp\ket{b}\bra{b}H^\perp\ket{i}
  \frac{1}{2} \PV\left[\frac{1}{E_i-E_b}+\frac{1}{E_a-E_b}\right]\label{e5}\\
  \bra{f}h_2\!\cdot\! h_1\ket{i}
  &= \sum_{a\in\eta=+1}
  \sum_{b\in\eta=-1}
  \bra{f}H^\perp\ket{b}\bra{b}H^\perp\ket{a}
  \frac{1}{2}\PV\left[\frac{1}{E_i-E_b}+\frac{1}{E_a-E_b}\right]
  \frac{\bra{a}H_1^\parallel\ket{i}}{E_i-E_a+i\varepsilon}\label{e6}
  .
\end{align}
Here we have used $E_f=E_i$ and the off-diagonal matrix element
$\bra{a}h_2\ket{i}$ in~(\ref{eq:h2offdiag}). The $E_a-E_b$ denominator
does not refer to $E_i$ and therefore does not have an OFPT analog. The terms that have the
OFPT structure $1/[(E_i-E_b)(E_i-E_a)]$ are $1/2$ the size of OFPT.

\subsection{The $h_3$ term}

Because $h_3 = \tfrac18\comm{J}{Q_2^\perp}$ is a commutator, it contains a
combination of the three orderings of two $H^\perp$ vertices and one
$H_1^\parallel$ vertex. The $H^\perp H_1^\parallel H^\perp$ ordering is
supplied solely by $h_3$. The $\eta$-flipping operator $Q_2^\perp$ is
determined by the $\eta$-flipping operator $Q_1$ through
$\comm{Q_2^\perp}{H_0}=-\comm{Q_1}{H_1^\parallel}$. In the energy eigenbasis,
this gives
\begin{align}\label{eq:Q2perp}
  \bra{b} Q_2^\perp\ket{i}
  &= \PV\frac{\bra{b}\comm{Q_1}{H_1^\parallel}\ket{i}}{E_b - E_i}\\
  &= \PV \sum_a \biggl(\frac{\bra{b}J\ket{a}\bra{a}H_1^\parallel\ket{i}}
    {(E_b - E_a)(E_b - E_i)}
    - \frac{\bra{b}H_1^\parallel\ket{a}\bra{a}J\ket{i}}
  {(E_a - E_i)(E_b - E_i)}\biggr)
\end{align}
where $b$ and $i$ lie in opposite $\eta$-sectors, while the sector of $a$
depends on the term. Using $J=2\eta H^\perp$, the matrix element of the
$\frac18 J Q_2^\perp$ term in $h_3$ is
\begin{equation}\label{eq:h3Hperp1}
  \tfrac18 \bra{f} J Q_2^\perp \ket{i}
  = \frac{1}{2} \sum_{b,a}\,
  \bra{f}H^\perp\ket{b}\,\PV\biggl(\frac{\bra{b}H^\perp\ket{a}\bra{a}H_1^\parallel\ket{i}}
    {(E_b - E_a)(E_i - E_b)}
    + \frac{\bra{b}H_1^\parallel\ket{a}\bra{a}H^\perp\ket{i}}
    {(E_i - E_a)(E_i - E_b)}
  \biggr)\,.
\end{equation}
Similarly, the other term in $h_3$ using $E_f=E_i$ gives
\begin{equation}\label{eq:h3Hperp2}
  -\tfrac18 \bra{f} Q_2^\perp J\ket{i}
  = \frac{1}{2}  \sum_{b,a}\,\PV
  \biggl(\frac{\bra{f}H^\perp\ket{a}\bra{a}H_1^\parallel\ket{b}}
    {(E_i - E_a)(E_i - E_b)}
    + \frac{\bra{f}H_1^\parallel\ket{a}\bra{a}H^\perp\ket{b}}
    {(E_b - E_a)(E_i - E_b)}
  \biggr)\,\bra{b}H^\perp\ket{i}.
\end{equation}

\subsection{The real parts}

We now compare the real parts of these results as distributions with OFPT.  To simplify notation, we define
\begin{equation}
 x=E_i-E_a,\qquad y=E_i-E_b,\qquad
 P_x=\PV\frac1x,\qquad \delta_x=\delta(x).
\end{equation}
We shall also suppress the products of matrix elements. In standard OFPT, the real part is
\begin{equation}\label{eq:OFPTthirdreal}
 C^{\rm OFPT}
 =\operatorname{Re}\!\left[\frac1{x+i0}\frac1{y+i0}\right]
 =P_xP_y-\pi^2\delta_x\delta_y.
\end{equation}

For the ordering $H^\perp H_1^\parallel H^\perp$ (middle ordering), the second term
in Eq.~(\ref{eq:h3Hperp1}) and the first term in
Eq.~(\ref{eq:h3Hperp2}) combine to give
\begin{equation}\label{eq:Z2middlereal}
 C_{\rm mid}^{Z_2}=P_xP_y.
\end{equation}
Consider either of the orderings
$H_1^\parallel H^\perp H^\perp$ or $H^\perp H^\perp H_1^\parallel$ (outer ordering). For $H_1^\parallel H^\perp H^\perp$, the real part of
Eq.~(\ref{e5}) and the first term in Eq.~(\ref{eq:h3Hperp2}) combine to give
\begin{equation}\label{eq:Z2outerreal}
 C_{\rm out}^{Z_2}
 =\frac12P_xP_y+\frac12P_{y-x}(P_x-P_y).
\end{equation}
The second term contains
the nonstandard energy denominator $y-x=E_a-E_b$. The
Poincar\'e-Bertrand identity gives
\begin{equation}\label{eq:PBthird}
 P_{y-x}(P_x-P_y)=P_xP_y-\pi^2\delta_x\delta_y,
\end{equation}
and so
\begin{equation}\label{eq:Z2outerreal2}
 C_{\rm out}^{Z_2}=P_xP_y-\frac{\pi^2}{2}\delta_x\delta_y.
\end{equation}

Thus, the real part of the Z$_2$PT orderings agrees pointwise with OFPT away from
vanishing denominators. As distributions, the middle ordering contains
none of the OFPT double-delta term, whereas the outer ordering contains
half of it. The $H_1^\parallel H_1^\parallel H_1^\parallel$ product
matches OFPT.

These differences do not affect primitive UV divergences. Every
double-delta term has support only when
$E_a=E_b=E_i$, and the fixed $E_i$ bounds all momenta carried by the on-shell intermediate
states. Any UV divergence confined to an uncut subgraph is a subdivergence
and is removed by lower-order renormalization. Thus, local primitive UV-divergent parts agree with OFPT
even though the real contact terms can differ.

The imaginary parts  of matrix elements in Z$_2$PT will differ from OFPT by construction, since this is
what makes ghost parity symmetry manifest in Z$_2$PT. For the middle ordering,
the imaginary part is absent. For the outer ordering,
(\ref{e5}) and (\ref{e6}) retain the cut through the same-sector
intermediate state but omit the cross-sector cut. Only the
$H_1^\parallel H_1^\parallel H_1^\parallel$ product agrees with OFPT.

\subsection{Ghost appearances}
\label{sec:prescription}

The Z$_2$PT prescription is attached to the
type of transition, not to the species of particle. The $\PV$
prescription applies to energy denominators that arise internally within each
$h_i$, whereas the $+i\varepsilon$ prescription arises from time-ordered
products of $h_i$. A ghost, a field that flips ghost parity, can appear
in both contexts. Ghosts can occur as physical particles in the initial and
final states, in any number consistent with matching initial/final state
ghost parities. They can also occur in intermediate states that can go on
shell, in which case the $i\varepsilon$ prescription applies, provided that
the intermediate state ghost parity matches those of the initial and final
states. If the ghost parity does not match, the intermediate state is purely
virtual and the $\PV$ prescription applies. The $\PV$ prescription can also
involve an intermediate state whose ghost parity matches those of the initial
and final states, but then the energy difference is taken with respect to
another intermediate state of opposite ghost parity.

In the fakeon proposal~\cite{anselmi_fakeon}, the ghost is replaced by a
virtual degree of freedom whose propagator is defined using the
principal-value prescription,
\begin{equation}
  D_{\text{fakeon}}(p^2) = \PV\!\left[\frac{1}{p^2-m^2}\right]
  = \operatorname{Re}\!\left[\frac{1}{p^2-m^2+i\varepsilon}\right].
\end{equation}
This is the same as a single ghost exchange described by
$h_2$. Beyond this correspondence, however, the two frameworks differ
substantially. In the fakeon proposal, $\PV$ is attached to the particle type;
    the fakeon always uses $\PV$ and never appears on shell.
The fakeon requires a sophisticated
    nonanalytic extension of the standard Wick rotation~\cite{fakeon_fw}.
In Z$_2$PT, the $\PV$ prescription enters through OFPT-style energy
    denominators, which define the amplitude without contour ambiguities.
Lastly, Z$_2$PT is mechanically obtained from a similarity transformation.

\section{Structure at fourth order when $h_1=0$}
\label{sec:h4structure}

To simplify the fourth order analysis we set $h_1=H_1^\parallel=0$, so that $H_1=H^\perp$ and every odd $h_i$
vanishes. The Dyson expansion contains two contributions,
\begin{equation}
  T[h_2(t_1)h_2(t_2)]\qquad\text{and}\qquad h_4(t_1).
\end{equation}
The first contains a same-sector OFPT denominator between two
instantaneous $h_2$ interactions, while all internal denominators of $h_2$
and $h_4$ arise from cross-sector inversions.

We first consider contributions for which none of the energy denominators
vanishes. The prescription is then not triggered, and the comparison with OFPT is
at the level of ordinary rational functions. Consider a fixed chain of
states,
\begin{equation}\label{eq:h4chain}
  \ket{i_+}\longrightarrow\ket{a_-}\longrightarrow\ket{b_+}
  \longrightarrow\ket{c_-}\longrightarrow\ket{f_+},
  \qquad E_f=E_i,
\end{equation}
where the subscripts denote the $\eta$ sectors. Define
\begin{equation}\label{eq:h4xyz}
  x=E_i-E_a,\qquad y=E_i-E_b,\qquad z=E_i-E_c.
\end{equation}
In addition to $x,y,z\neq0$, we assume $x-y=E_b-E_a\neq0$ and
$z-y=E_b-E_c\neq0$. We suppress the common product of matrix elements
\begin{equation}\label{eq:h4vertices}
  \bra{f}H^\perp\ket{c}\bra{c}H^\perp\ket{b}
  \bra{b}H^\perp\ket{a}\bra{a}H^\perp\ket{i}
\end{equation}
and compare its coefficients.

We start with the $h_2^2$ term. The state $\ket{b}$ lies between the two $h_2$ insertions. Using the
off-diagonal matrix element~(\ref{eq:h2offdiag}) gives
the coefficient
\begin{align}
 C_{22}
 &=\frac{1}{4y}
 \left(\frac{1}{z}+\frac{1}{z-y}\right)
 \left(\frac{1}{x}+\frac{1}{x-y}\right)\nonumber\\
 &=\frac{(2x-y)(2z-y)}
 {4xyz(x-y)(z-y)}.\label{eq:C22}
\end{align}
The factor $1/y$ is the same-sector OFPT denominator while the other
denominators arise from $h_2$ and connect states
in opposite sectors.

Next, we consider the $h_4$ term. When $H_1=H^\perp$, $Q_2$ is $\eta$-preserving, and from $Q^2=1$ it is
\begin{equation}\label{eq:Q2h1zero}
  Q_2=-\tfrac12\eta Q_1^2.
\end{equation}
Consequently, $Q_3^\perp$ is nonzero and is determined by
\begin{equation}\label{eq:Q3h1zero}
  \comm{Q_3^\perp}{H_0}=-\comm{Q_2}{H^\perp}.
\end{equation}
For states in definite sectors, the elementary matrix elements needed to
evaluate Eq.~(\ref{eq:simplified}) are
\begin{equation}\label{eq:JQ1sector}
  \bra{r}J\ket{s}=2\eta_r\bra{r}H^\perp\ket{s},\qquad
  \bra{r}Q_1\ket{s}
  =\frac{2\eta_r}{E_r-E_s}\bra{r}H^\perp\ket{s},
\end{equation}
where $\eta_r=\pm1$ is the sector of $\ket r$.

Keeping only the matrix elements belonging to the fixed chain, the two
$Q_3^\perp$ matrix elements in $[J,Q_3^\perp]$ are
\begin{align}
 \bra{c}Q_3^\perp\ket{i}
 &=\frac{1}{E_c-E_i}\left(
 \bra{c}Q_2\ket{a}\bra{a}H^\perp\ket{i}
 -\bra{c}H^\perp\ket{b}\bra{b}Q_2\ket{i}\right),\nonumber\\
 \bra{f}Q_3^\perp\ket{a}
 &=\frac{1}{E_f-E_a}\left(
 \bra{f}Q_2\ket{b}\bra{b}H^\perp\ket{a}
 -\bra{f}H^\perp\ket{c}\bra{c}Q_2\ket{a}\right).
 \label{eq:Q3chain}
\end{align}
Equation~(\ref{eq:Q2h1zero}) then reduces every $Q_2$ matrix element to two
$Q_1$ matrix elements.

For the chain~(\ref{eq:h4chain}), the three operator structures in $h_4$
give, respectively,
\begin{align}
 C_{4,Q_3}
 &= -\frac{x-y+z}{xz(x-y)(z-y)},\label{eq:C4Q3}\\
 C_{4,Q_1^3}
 &= \frac{3(x+z)}{8xz(x-y)(z-y)},\\
 C_{4,Q_1[Q_1,J]Q_1}
 &= \frac{x-2y+z}{8xz(x-y)(z-y)}.
\end{align}
Their sum is
\begin{equation}\label{eq:C4}
 C_4=\frac{3y-2x-2z}{4xz(x-y)(z-y)}.
\end{equation}

Thus, neither $h_2^2$ nor $h_4$ separately has the OFPT denominator
structure. Their sum, however, is
\begin{align}
 C_{22}+C_4=\frac{1}{xyz}.\label{eq:h4OFPTidentity}
\end{align}
  This establishes agreement with OFPT away from vanishing denominators. It also shows that any remaining difference must be
  distributionally supported where denominators vanish. We now determine this
  difference by retaining the principal-value prescriptions and the
  associated Poincar\'e-Bertrand contact terms. We want to compare to the OFPT result
\begin{align}
 \operatorname{Re}\!\left[
 \frac{1}{x+i0}\frac{1}{y+i0}\frac{1}{z+i0}\right]
 &=\PV\frac1x\PV\frac1y\PV\frac1z\nonumber\\
 &\quad-\pi^2\left[
 \delta(x)\delta(y)\PV\frac1z
 +\delta(x)\PV\frac1y\delta(z)
 +\PV\frac1x\delta(y)\delta(z)\right].\label{eq:tripleReal}
\end{align}

The real part of the $h_2^2$
contribution is
\begin{equation}\label{eq:C22PV}
 C_{22}^{\rm PV}
 =\frac14 P_y(P_z+P_{z-y})(P_x+P_{x-y}).
\end{equation}
 Define
\begin{equation}
 A=P_zP_{x-y}P_{z-y},\qquad
 B=P_xP_{x-y}P_{z-y},\qquad
 C=P_xP_zP_{z-y},\qquad
 D=P_xP_zP_{x-y},
\end{equation}
where each product contains three linearly independent linear forms and is therefore
well defined as a distribution.
The two Poincar\'e-Bertrand identities needed to reduce
Eq.~(\ref{eq:C22PV}) are
\begin{align}
 P_{x-y}(P_y-P_x)&=P_xP_y-\pi^2\delta_x\delta_y,\nonumber\\
 P_{z-y}(P_y-P_z)&=P_yP_z-\pi^2\delta_y\delta_z.
\label{e7}\end{align}
They give
\begin{equation}
 C_{22}^{\rm PV}=P_xP_yP_z+\frac14(A+C+2D)
 -\frac{\pi^2}{2}
 \label{e8}\left(\delta_x\delta_yP_z+P_x\delta_y\delta_z\right).
\end{equation}

For the $h_4$ contribution, before the rational simplifications in
(\ref{eq:C4Q3})--(\ref{eq:C4}), the three operator structures give
\begin{align}
 C_{4,Q_3}^{\rm PV}&=-\frac12(A+B+C+D),\nonumber\\
 C_{4,Q_1^3}^{\rm PV}&=\frac38(A+B),\nonumber\\
 C_{4,Q_1[Q_1,J]Q_1}^{\rm PV}&=\frac18(C+D),
\end{align}
and hence
\begin{equation}\label{eq:C4PV}
 C_4^{\rm PV}=-\frac18(A+B+3C+3D).
\end{equation}

We now make use of the distributional identity
\begin{equation}\label{eq:h4PVidentity}
 A-B-C+D=0.
\end{equation}
This identity is derived in Appendix A.
Combining (\ref{e8}), (\ref{eq:C4PV}), and~(\ref{eq:h4PVidentity}) then gives the complete real
fourth-order coefficient in Z$_2$PT,
\begin{equation}\label{eq:h4Z2real}
 C_{(4)}^{Z_2}=C_{22}^{\rm PV}+C_4^{\rm PV}=P_xP_yP_z-\frac{\pi^2}{2}
 \left(\delta_x\delta_yP_z+P_x\delta_y\delta_z\right).
\end{equation}

Thus, the contact terms for the adjacent pairs $(x,y)$ and $(y,z)$ have
one-half of their OFPT coefficients, while the contact term for the
nonadjacent pair $(x,z)$ is absent. Note that $x$ and $z$ reference energies of the
$\eta=-1$ states. This pattern parallels the third-order result; the double-delta contact term is retained with one-half of its OFPT coefficient when only one of the delta functions references a $\eta=-1$ state energy and it is absent when both delta functions do.

The contact terms are supported where at least two intermediate
states have energy $E_i$. Since the multiparticle states have positive energy,
the fixed $E_i$ bounds every momentum carried by those on-shell states. Any UV
divergence confined to the remaining uncut subgraph is a subdivergence. Thus,
these contact terms cannot produce a primitive UV divergence.

In standard scattering theory, the double-delta terms represent successive on-shell rescatterings and form the even-cut contribution to the real amplitude. Such terms can affect phase
shifts and cross sections and can carry threshold or infrared nonanalyticity, including mass-dependent logarithms. Recent work on
four-derivative theories has emphasized infrared-sensitive
$\log(p^2/m^2)$ terms and their possible interpretation as physical running
distinct from conventional $\mu$-running
\cite{Buccio:2024quadratic,Buccio:2025conformal,Salvio:2026infrared}. This reflects a logarithmic sensitivity to the momentum region between the ghost mass and the external scale that is not found in two-derivative theories. Whether the double-delta terms generate these $\log(p^2/m^2)$ terms requires an explicit OFPT decomposition of the relevant loop amplitudes.
If they do generate these terms then Z$_2$PT will alter their coefficients.

We have looked at fourth order with $h_1=0$ and have found that Z$_2$PT and OFPT have the same local primitive
UV-divergent part, even though their complete
real and imaginary parts differ.

\section{Renormalization}
\label{sec:renormalizability}

The original ghost theory and the superselected ghost theory can produce
different physical results. The standard perturbative splitting
$H=H_0+H_1$ is consistent for the original theory but does not respect the
superselection rule at finite order. Nevertheless, it can still determine
the local UV counterterms of the superselected theory if the local
UV-divergent parts agree. We have found this agreement through third order
and, when $h_1=0$, through fourth order. In this section, we assume this agreement at all orders.

To renormalize Z$_2$PT we can begin with the
renormalized decomposition
\begin{equation}
 H_R=H_{0,R}+H_R^\parallel+H_R^\perp
\end{equation}
and use the renormalized fields, masses, and couplings in constructing
\begin{align}
 h_R&=g_RH_Rg_R^{-1}=h_{0,R}+\sum_{n\geq1}h_{n,R},\\
 g_R&=\sqrt{\eta Q_R},\qquad h_{0,R}=H_{0,R},\qquad h_{1,R}=H_R^\parallel.
\end{align}
Calculations within Z$_2$PT are then renormalized by adding, as needed, local
$\eta$-preserving counterterms. These
belong to the local sector represented by $h_{0,R}$ and $h_{1,R}$. No $\eta$-flipping counterterm
is needed or generated, and no independent counterterms are assigned
to the nonlocal vertices $h_{2,R},h_{3,R},\ldots$. The agreement of the primitive local UV
divergences implies that the local counterterms required in Z$_2$PT have the
same coefficients as the corresponding counterterms in the original ghost theory. These coefficients
can be determined in either theory.

This does not mean that a coupling $\lambda^\perp$
appearing in nonlocal vertices does not run. It inherits the running as determined by the
original local theory, as reflected in the scale dependence of the renormalized coupling $\lambda_R^\perp(\mu)$
that has been used in the construction of $h_R$.
This determination of $\lambda^\perp$, which is otherwise a finite free parameter in Z$_2$PT, is a matching condition. It may also be the case that a $\lambda^\parallel$ and a $\lambda^\perp$ in Z$_2$PT
share a common origin as a single coupling in the original local theory.

Any link between UV divergences and physical cuts that may exist
in normal theories is not present in Z$_2$PT.
This was already observed at second order; $h_2$ removes the
cross-sector cut while retaining the principal-value contribution containing
the local UV divergence. It is the
agreement of these local UV-divergent parts with OFPT, which we have now found to fourth order, that is the relevant result for renormalization. Z$_2$PT can be renormalized directly with
local $\eta$-preserving counterterms, and then the renormalized parameters determined
in this sector are used throughout the transformed
vertices $h_{2,R},h_{3,R},\ldots$. No $\eta$-flipping
counterterms are required and instead the $\eta$-flipping couplings found
in $h_{2,R},h_{3,R},\ldots$ are matched to the renormalized couplings in $H^\perp$.

\section{Comparison to other similarity transformations}
\label{sec:comparison}

The similarity transformation $h=gHg^{-1}$ that makes a ghost parity
symmetry manifest has relatives in two other frameworks,
pseudo-Hermitian/PT-symmetric quantum mechanics and the
Schrieffer--Wolff transformation. We compare Z$_2$PT with each in turn.

\subsection{Pseudo-Hermitian and PT-symmetric quantum mechanics}\label{sec:PHcompare}

In pseudo-Hermitian quantum mechanics (PH-QM), a non-Hermitian Hamiltonian
$H$ with a real spectrum is mapped to a Hermitian Hamiltonian
$h$ via a positive metric operator $G=g^2$ as $h=g H g^{-1}$
\cite{Mostafazadeh:2001nr}. A systematic
perturbative construction of the metric operator and the equivalent
Hermitian Hamiltonian was developed in~\cite{Mostafazadeh:2005metric,Mostafazadeh:2010review}
and, in the PT-symmetric formulation, in~\cite{Bender:2004prd,Bender:2007review}. The starting
point is $H=H_0+\epsilon H_1$, where $H_0$ is Hermitian and $H_1$ is
anti-Hermitian. The positive metric is parametrized as
$G=e^{-Q}$ with $Q=\sum_{j=1}^\infty Q_j\epsilon^j$, where
$Q_j$ are Hermitian operators determined iteratively according
to~\cite{Mostafazadeh:2010review},
\begin{align*}
  [H_0,Q_1] &= -2H_1, \label{eq:PHQ1}\\
  [H_0,Q_2] &= 0, \\
  [H_0,Q_3] &= -\tfrac16[[H_1,Q_1],Q_1], \\
  [H_0,Q_4] &= -\tfrac16\bigl([[H_1,Q_1],Q_2]+[[H_1,Q_2],Q_1]\bigr).
\end{align*}
The recursion follows from expanding the pseudo-Hermiticity
condition $H^\dagger G=GH$ order by order. For a discrete, nondegenerate
spectrum of $H_0$, each relation $[H_0,Q_j]=R_j(Q_1,\ldots,Q_{j-1})$ is
inverted in the $H_0$ eigenbasis to fix the off-diagonal matrix elements
of $Q_j$ through energy denominators, while $Q_2$ and the diagonal parts of
the other $Q_j$ remain subject to metric freedom.
The equivalent Hermitian Hamiltonian is then
\begin{equation}
  h = e^{-Q/2} H e^{Q/2}.
\end{equation}

The structural parallels and differences between PH-QM and Z$_2$PT are as follows.
\begin{itemize}
\item
In PH-QM the metric is expanded, whereas in
Z$_2$PT the exact ghost parity is expanded,
$Q=\eta+Q_1+Q_2+\cdots$.

\item
The PH-QM rules $[H_0,Q_j]=R_j$ involve increasingly
nested commutators of $H_1$ with $Q_k$. The Z$_2$PT rules
$[Q_n,H_0]=-[Q_{n-1},H_1]$ are structurally simpler.

\item
PH-QM assumes that $H_1$ is purely
anti-Hermitian, which forces $h_1=0$; higher odd orders are tied to
metric freedom. In Z$_2$PT, when $H_1$ is
purely $\eta$-flipping, $[\eta,h]=0$ forces every odd $h_i$ to vanish.

\item
The PH-QM perturbative expansion has been developed for
quantum mechanical systems with a discrete, nondegenerate spectrum.
Z$_2$PT is formulated in a superselected QFT on Fock space, where
the energy denominators are regulated.

\item
The primary object in Z$_2$PT is not the positive inner product, but rather
$\eta$. The two frameworks differ in how positivity is achieved, through the
inner product in PH-QM and through superselection in Z$_2$PT.

\item
The PT-symmetric subclass of PH-QM is equipped with a
  $\mathcal{CPT}$ metric and a charge $\mathcal C$~\cite{Mostafazadeh:2010review,Bender:2007review}
analogous to $Q$, with
$[\mathcal{C},H]=0$ and $[\mathcal{C},H_0]\neq0$. In Z$_2$PT, we also have
$[\eta,h]=0$. The analog of $\eta$ is $\mathcal{P}$, but, to our
knowledge, the statement $[\mathcal{P},h]=0$ does not appear explicitly in the PT-symmetric QM
literature.
\end{itemize}

\subsection{The Schrieffer--Wolff transformation}\label{sec:SW}

The Schrieffer--Wolff (SW) transformation of~\cite{Bravyi:2011fda}
block-diagonalizes $H=H_0+H_1$ via the rotation
$U=\sqrt{R_{P_0}R_P}$ and forms the effective Hamiltonian
$H_{\rm eff}=P_0 U H U^\dagger P_0$. Here, $P$ and $P_0$ are projection
operators, while $R_P$ and $R_{P_0}$ are reflection operators that satisfy
$R_P^2=R_{P_0}^2=1$. With $R_{P_0}\!\leftrightarrow\!\eta$,
$R_P\!\leftrightarrow\!Q$,
$V_d\!\leftrightarrow\!H_1^\parallel$, and
$V_{od}\!\leftrightarrow\!H^\perp$, the rotation $U$ is algebraically the
same object as $g=\sqrt{\eta Q}$. SW satisfies
$[R_P,H]=0$ and $[R_{P_0},H_{\textit{eff}}]=0$. Setting $U=e^S$ yields the
expansion $S=\sum_{j=1}^\infty S_j\epsilon^j$ for the rotation generator,
closely analogous to that of PH-QM. The
Z$_2$PT commutation rules are replaced by the more involved recursion
$S_n=-{\cal L}\hat V_d(S_{n-1})+\sum(\cdots)$ of \cite{Bravyi:2011fda}.

Z$_2$PT and SW yield the same algebraic expressions for $h_1$ and $h_2$.
From $g=\sqrt{\eta Q}=1+\frac12\eta Q_1+\cdots$ and
$U=e^S=1+S_1+\cdots$ one identifies $S_1 = \tfrac12\eta Q_1$.
Identifying $V_{od}$ with $H^\perp$ then gives
\begin{equation}
  h_2=\tfrac18[J,Q_1]=\tfrac14\eta\{H^\perp,Q_1\}
  =\tfrac12[S_1,V_{od}]=H_{\mathrm{eff},2}  .
\end{equation}
At higher orders, the expressions differ in form because $Q_k$ and
$S_k$ are nonlinearly related, expanding $\eta Q=e^{2S}$ gives
$\eta Q_1=2S_1$, $\eta Q_2=2S_2+2S_1^2$,
$\eta Q_3=2S_3+2\{S_1,S_2\}+\frac43S_1^3$, etc.
In principle, Z$_2$PT could adopt $g=e^S$ and the $S$ expansion. The genuine
differences between SW and Z$_2$PT are, therefore, structural.
\begin{itemize}
\item
The SW perturbation $V_{od}$ is Hermitian, and the Z$_2$PT
$H^\perp$ is anti-Hermitian.

\item
SW defines its blocks by an energy interval with a spectral gap
$\Delta$; Z$_2$PT defines them by the ghost parity $\eta$.

\item
SW's gap makes every cross-block denominator nonzero, so no
prescription is needed.

\item
SW projects onto one low-energy block and integrates out the other;
$H_{\rm eff}=P_0 U H U^\dagger P_0$ acts on a single subspace.
Z$_2$PT keeps both $\eta$-sectors; $h=gHg^{-1}$ is the full
block-diagonal operator on a Hilbert space in which both sectors
are physical.

\end{itemize}

\section{Conclusion}\label{conc}

A quantum field theory with a wrong-sign kinetic term, a ghost theory, is
unitary with respect to the indefinite inner product $\eta$. Imposing a
superselection rule based on a ghost parity $Q$ that satisfies $\comm{Q}{H}=0$ is sufficient to
provide a probability interpretation. The obstacle is perturbative, the
splitting $H=H_0+H_1$ does not respect the rule at any finite order, because
neither $\comm{Q}{H_0}$ nor $\comm{Q}{H_1}$ vanishes separately. The
similarity transformation $h=gHg^{-1}$, with $g=\sqrt{\eta Q}$, removes this
obstacle by making $\eta$ a manifest conserved charge,
$\comm{\eta}{h}=0$, order by order in Z$_2$PT. In~\cite{paper1} it was
argued that the superselection rule ensures that the optical theorem is satisfied with
positive probabilities and that the spectral representation of the propagator has standard analytic properties. Z$_2$PT provides a perturbative realization of these results.

Z$_2$PT and its nonlocal interactions resemble old-fashioned perturbation
theory (OFPT), and this comparison helps to identify differences between the
superselected theory and the original theory. The nonlocal
terms in $h$ repackage cross-sector transitions in a manner that respects the
superselection rule. Consequently, cross-sector energy denominators carry the
principal-value prescription. Each $h_i$ contains different orderings of
interaction operators, all evaluated at a single instant in time. Same-sector
transitions involve standard time ordering of interaction operators and give
rise to energy denominators with the standard form and prescription. Such
transitions can involve particles from both sectors, so the prescription is
attached to the transition type rather than to the particle species.

At second order, the real part of the Feynman amplitude is reproduced, but not
necessarily its full imaginary part. At third order, Z$_2$PT agrees pointwise
with OFPT away from vanishing denominators. At fourth order with $h_1=0$, the nonstandard denominators in
$h_2^2$ and $h_4$ likewise combine to give the OFPT result pointwise. The local UV-divergent parts, therefore, agree with those of the original ghost
theory.

The imaginary parts of Z$_2$PT amplitudes differ from OFPT by construction, since this
is what brings Z$_2$PT in line with the superselection rule. We have shown how
the real parts of amplitudes
also differ, due to double-delta contact terms at third and fourth order.
These latter differences occur when two intermediate states have the
initial/final state energy, and at least one of these two states is in the sector opposite to the initial/final state.
It would be interesting to understand the implications of these differences more fully.

\appendix
\section{}
A derivation of \eqref{eq:h4PVidentity} follows from the two Poincar\'e-Bertrand identities in (\ref{e7}). First we introduce four more triple products that are well defined since they contain three linearly independent linear forms,
  \begin{equation}
   E=P_yP_zP_{x-y},\qquad
   F=P_xP_yP_{z-y},\qquad
   G=P_yP_{x-y}P_{z-y},\qquad
   T=P_xP_yP_z.
  \end{equation}
  Multiplying the second Poincar\'e-Bertrand identity by $P_{x-y}$
  and by $P_x$, respectively, and using $P_{x-y}\delta_y\delta_z=P_x\delta_y\delta_z$, gives
  \begin{align}
   G-A&=E-\pi^2P_x\delta_y\delta_z,\nonumber\\
   F-C&=T-\pi^2P_x\delta_y\delta_z.
  \end{align}
  Similarly, multiplying the first Poincar\'e-Bertrand identity by
  $P_{z-y}$ and by $P_z$ gives
  \begin{align}
   G-B&=F-\pi^2\delta_x\delta_yP_z,\nonumber\\
   E-D&=T-\pi^2\delta_x\delta_yP_z.
  \end{align}
  Equivalently,
  \begin{align}
   A&=G-E+\pi^2P_x\delta_y\delta_z,\nonumber\\
   B&=G-F+\pi^2\delta_x\delta_yP_z,\nonumber\\
   C&=F-T+\pi^2P_x\delta_y\delta_z,\nonumber\\
   D&=E-T+\pi^2\delta_x\delta_yP_z.
  \end{align}
  We see that all principal-value and contact terms cancel pairwise when substituted into \eqref{eq:h4PVidentity}, thus verifying this identity.

\end{document}